# Magnetism of a sigma-phase $Fe_{60}V_{40}$ alloy: magnetic susceptibilities and magnetocaloric effect studies

Maria Bałanda[1], Stanisław M. Dubiel[2*] and Robert Pełka[1]

[1]Institute of Nuclear Physics, Polish Academy of Science, PL-31-342 Kraków, Poland
[2]AGH University of Science and Technology, Faculty of Physics and Applied Computer Science, PL-30-059 Kraków, Poland

## Abstract

Magnetic properties of a sigma-phase $Fe_{60}V_{40}$ intermetallic compound were studied by means of ac and dc magnetic susceptibility and magnetocaloric effect measurements. The compound is a soft magnet yet it was found to behave like a re-entrant spin-glass system. The magnetic ordering temperature was found to be $T_C \approx 170$ K, while the spin-freezing temperature was ~164 K. Its relative shift per decade of ac frequency was 0.002, a value smaller than that typical of canonical spin-glasses. Magnetic entropy change, $\Delta S$, in the vicinity of $T_C$ was determined for magnetic field, $H$, ranging between 5 and 50 kOe. Analysis of $\Delta S$ in terms of the power law yielded the critical exponent, $n$, vs. temperature with the minimum value of 0.75 at $T_C$, while from the analysis of a relative shift of the maximum value of $\Delta S$ with the field a critical exponent $\Delta=1.7$ was obtained. Based on scaling laws relationships values of other two exponents viz. $\beta=0.6$ and $\gamma=1$ were determined.

*Corresponding author: Stanislaw.Dubiel@fis.agh.edu.pl

## 1. Introduction

The magnetism of a sigma (σ) phase in the Fe-V alloy system has been known since 1960 [1]. However, it has been regarded as ferromagnetism up to recently, when experimental evidence, by means of *DC* magnetization measurements, was found that the magnetism has a re-entrant character [2,3]. Recent AC magnetic susceptibility measurements performed on the magnetically strongest sample of a σ-$Fe_{65.6}V_{34.4}$ alloy ($T_C \approx 312K$) confirmed these finding and gave evidence that several parameters of merit had values typical of canonical spin-glasses [4]. The latter feature seems rather unexpected for a system in which a concentration of magnetic atoms is so high, but it can be understood in terms of a highly itinerant character of magnetism in the studied system [5], hence very long-ranged magnetic interactions. It is of interest to verify whether or not the behavior found for the magnetically strongest sample depends on the strength of magnetism (Curie temperature). The Fe-V system gives for that purpose a good opportunity, as the σ-phase can be formed in a wide compositional range viz. ~32 - ~66 at. %V [6] in which $T_C$ changes by two orders of magnitude [4].

Another reason to study magnetism of these intermetallic compounds is related to a magnetocaloric effect. Due to the superb soft magnetism and magnetic ordering temperatures reaching 312 K [5], the σ-$Fe_{1-x}V_x$ intermetallics may be considered as prospective magnetocaloric materials. Magnetocaloric effect (MCE) is of great technological importance since it can be used for cooling in the process of adiabatic demagnetization which is an environmentally friendly technique. MCE is an intrinsic attribute of all magnetic materials, although its intensity depends on particular properties of each material. Two measures of merits for MCE are: an adiabatic temperature change, $\Delta T_{ad}$, and an isothermal magnetic entropy change, $\Delta S$, in response to a change of the magnetic field, $\Delta H$. Last years there has been an intense quest for suitable materials showing large values of $\Delta T_{ad}$ and/or of $\Delta S$ at $\Delta H$ in the range of 20 kOe and at temperatures close to the room temperature [7,8]. The effect itself is also interesting from the fundamental point of view as it reveals thermodynamic properties and the critical behaviour of a material in the vicinity of the magnetic phase transition.

Corresponding results obtained on a σ-$Fe_{60.1}V_{39.9}$ (hereafter referred to as σ-$Fe_{60}V_{40}$) sample ($T_C \approx 170K$) using AC and DC magnetization techniques are reported and discussed in this paper.

## 2. Experimental

Magnetic measurements performed by means of a MPMS XL SQUID magnetometer were carried out on a powder sample of a σ-phase $Fe_{60}V_{40}$ alloy whose mass was 31.3 mg. Its preparation and characterization details are given elsewhere [9]. Both real, $\chi'$, and imaginary, $\chi''$, components of the AC magnetic susceptibility, $\chi_{AC} = \chi' - i\chi''$, were registered as a function of increasing temperature, $T$, in the range of 5-340 K. Amplitude of the oscillating field, $H_{AC}$, was equal to 1 Oe, 2 Oe or 3 Oe, frequencies, $f$, ranged between 0.1 Hz and 1000 H. An influence of the applied DC magnetic field, $H_{DC}$, on the temperature dependence of $\chi_{AC}$ was investigated, too. Isothermal magnetization curves at 2 K, 60 K and 120 K up to 70 kOe were measured, as well as the temperature and time dependence of a small remanent magnetization present in the sample after switching off the field. The temperature dependence of the zero-field cooled (ZFC) and field-cooled (FC) magnetization was measured at several values of the applied field. In order to determine the magnetocaloric effect (MCE), a series of magnetization vs. temperature runs on heating from 100 K to 250 K were carried out at twelve different values of $H_{DC}$ ranging from 2.5 kOe up to 50 kOe.

## 3. Results and discussion

### 3.1. AC susceptibility – effect of frequency

The temperature dependence of $\chi'$ and $\chi''$ components of AC magnetic susceptibility, measured at frequency $f$ = 6 Hz and 216 Hz, is shown in Fig. 1. The imaginary component amounts to ~3.7 % of the real one, which points to a rather weak relaxation effects in the studied alloy. Values of both components depend on frequency, increase with temperature and have a maximum close to the temperature of magnetic phase transition. The shape of the $\chi'(T)$ curve differs from that characteristic of canonical spin glasses (C-SGs) e.g. CuMn, AuMn, AuFe [10] or even from the one found recently for σ-FeMo, for which a well-defined cusp with concave slopes was observed [11]. The χ-curves shown in Fig. 2 are similar to those measured recently for the σ-$Fe_{66}V_{34}$ alloy [4], in which a 6 % increase in concentration of magnetic carriers raised the magnetic ordering transition (Curie temperature) above the room temperature. An interlacement observed in the χ'' curves measured at T ≈ 15 K and T ≈ 60 K with different frequencies of the driving field is likely due to various local anisotropies of particular magnetic sublattices hence different relaxation times.

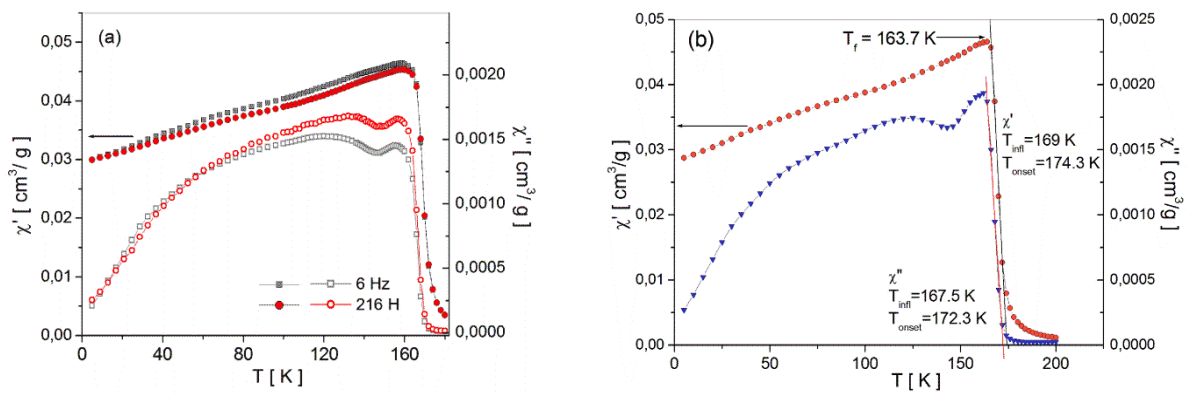

Fig. 1 (a) real, χ′, and imaginary, χ″, components of the AC magnetic susceptibility for the $Fe_{60.1}V_{39.9}$ versus temperature for measured at frequency 6 Hz and 216 Hz. In (b) χ′ and χ″ measured at 10 Hz are shown; characteristic temperatures are indicated.

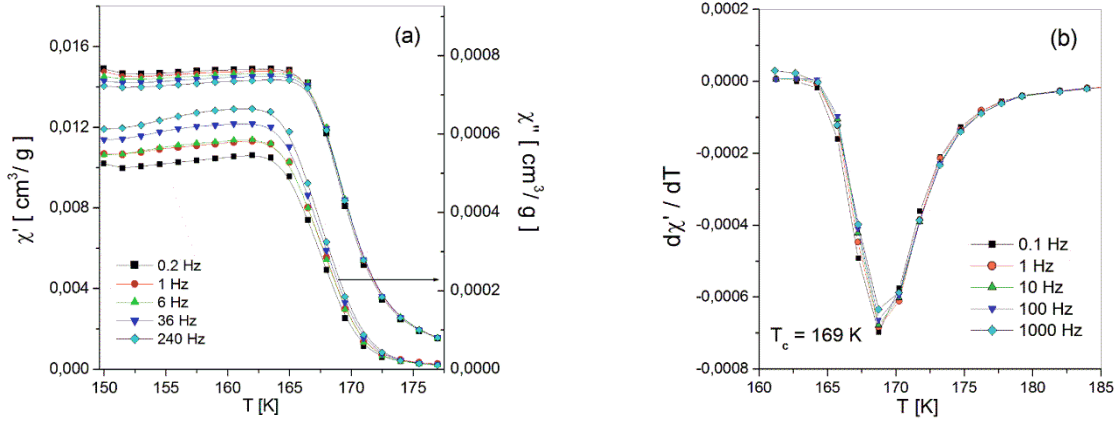

Fig. 2 (a) Temperature dependence of $\chi'$ and $\chi''$ close to the magnetic transition measured at several frequencies shown and $H_{AC}$ = 3 Oe., and (b) the temperature derivative of $\chi'$, $d\chi'/dT$, measured with $H_{AC}$ = 1 Oe at different frequencies displayed.

The measured $\chi'$ and $\chi''$ curves can be used to determine the magnetic ordering temperature, $T_C$. There is no unique way for doing that. Using the inflection method one gets $T_C$ = 169 K (from $\chi'$) and $T_C$=167.5 K (from $\chi''$) for $f$=10 Hz. The corresponding values obtained from the tangent method are 174.3 K and 172.3 K, respectively. The peak positions of $\chi'$ and $\chi''$ curves are 163.7 K and 162.5 K, respectively. The data displayed in Fig. 3a shows that at temperatures close to the magnetic transition the frequency has a slightly different effect on $\chi''$ than on $\chi'$. Figure 3b illustrates a $d\chi'/dT$ behavior in the vicinity of $T_C$ for various frequencies. It is evident that the minimum of the derivative, which defines $T_C$, hardly depends on $f$ i.e. it can be regarded as a temperature at which the magnetic phase transition occurs.

The temperature at which the AC magnetic susceptibility has its maximum is known as a spin freezing one, $T_f$. It is also well known that it depends on frequency, namely, it shifts toward higher temperatures when $f$ increases. The relative shift of $T_f$ per decade of frequency, $RST$, defined by the following formula:

$$RST = \frac{\Delta T_f / T_f}{\Delta \log f} \qquad (1)$$

is often used as a figure of merit for distinguishing between spin-glasses and superparamagnets [12]. Values of $RST$ can be also used to differentiate between various types of SGs. In the present case, $RST$=0.002, as determined from $\chi'$ measured with $H_{AC}$=3 Oe, while that determined using $H_{AC}$ = 1 Oe is equal to 0.003. The shift of $\chi''$ peak (Fig. 3a) yields the value of 0.0017. The presently found $RST$ values are slightly larger than those determined for the σ-phase $Fe_{65.9}V_{34.1}$ compound [4], yet smaller than the ones typical of the canonical spin glasses ($RST$≈0.005). A very weak frequency dependence of $T_f$ attests to a weak character of the spin-glass in the σ-FeV alloys with a high iron concentration, and it can be understood in terms of a long range interaction between the magnetic moments in these topologically close-packed structures. The latter is a consequence of an itinerant character of magnetism in the σ-phase intermetallic [13].

The frequency dependent $\chi'(T)$ data were analyzed in the frame of two relevant models. Applying the Vogel-Fulcher relation:

$$f = f_0^{VF} \exp\left(-\frac{E_a}{k_B(T_f - T_0^{VF})}\right) \quad (3)$$

we found $f_o^{VF}$= 7·10$^{13}$ Hz, the Vogel-Fulcher parameter $T_o^{VF}$= 164 K, and the activation energy $E_a/k_B$ =120 K. The latter is in line with the corresponding value of 101 K found for the σ-Fe$_{66}$V$_{34}$ alloy [4]. Using the critical slowing-down law:

$$f = f_o\left(\frac{T_f}{T_{SG}} - 1\right)^{zv} \quad (4)$$

a spin-glass temperature $T_{SG}$ = 166.6 K and the dynamic exponent $zv$=9.4 were obtained. The value of $zv$=8.5 was determined previously for the σ-Fe$_{66}$V$_{34}$ alloy [4].

## 3.2. AC susceptibility – effect of DC bias field

A DC bias magnetic field has a dramatic effect on the AC magnetic susceptibility curves – see Fig. 4. The most visible response is a decrease of the intensity of the $\chi'$ and $\chi''$ curves which is especially strong in the vicinity of $T_f$. At $H_{DC}$ = 200 Oe, the $\chi'$-curve are rounded and the maximum at $T_f$ can be hardly seen. In stronger fields (500 and 1000 Oe) the $\chi'$ peak again appears at ~170.5 K and ~173.2 K for $H_{DC}$=500 and 1000 Oe, respectively. This means that the DC bias field shifted the $T_f$ - position towards higher temperature ($T_f$ =163.7 K for $H_{DC}$=0). The shape of the high-field (500, 1000 Oe) $\chi'$ curves *at T<$T_f$* resembles the one characteristic of ferromagnetically ordered systems. This can signify that the applied DC magnetic field aligned magnetic moments. Noteworthy, no peaks around $T_f$ exist in the $\chi''$ curves measured in the DC bias field. A similar behavior of χ' was observed in other systems showing the reentrant SGs e.g. in Ni-Co-Mn-Sb Heusler alloys [14], interestingly, in the ferromagnetic gadolinium [15], disordered Cr$_3$Fe ferromagnet [16], and in the layered 2D Heisenberg magnet [17], to give few examples. Such an anomaly was explained in terms of short range spin fluctuations close the ordering transition.

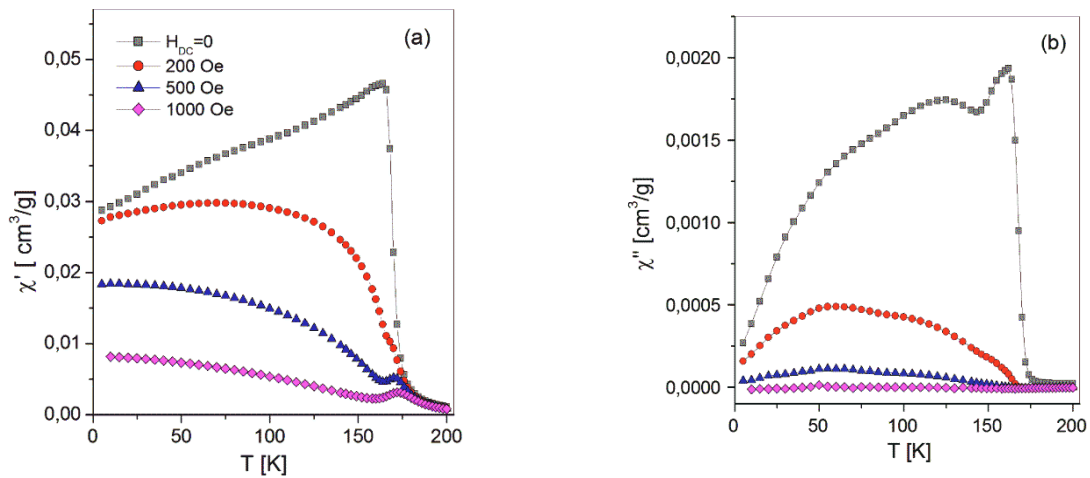

Fig.3 (a) Real, $\chi'$, and (b) imaginary, $\chi''$, parts of the AC susceptibility versus temperature, recorded for the σ-$Fe_{60.1}V_{39.9}$ sample in a bias DC magnetic field of 200 Oe, 500 Oe and 1 kOe for $H_{ac}$ = 3 Oe, f = 10 Hz. Data obtained in zero bias field are added for comparison.

### 3.3. DC magnetization

Temperature dependence of a static magnetic susceptibility, $\chi=M/H_{DC}$, was measured in the ZFC and FC modes at several values of the $H_{DC}$ field. ZFC and FC branches collapsed into one curve already for $H_{DC}$ = 200 Oe what corroborates the weak character of the spin-glass state observed in the studied system. A fit of the Curie-Weiss law to the 1 kOe FC data in the temperature interval of 190-300 K (see Fig. 4a) yielded an effective magnetic moment in the paramagnetic state $\mu_{eff}$ = 2.98 $\mu_B$ and the Curie-Weiss temperature $\theta_{CW}$ = 178 K. These figures can be further used, firstly, to demonstrate an itinerant character of magnetism in the studied sample, and secondly, to determine a degree of spin frustration, DSF. Concerning the former, the figure of merit viz. a ratio $\mu_{eff}/\mu_s$=7.5 which according to the Rhodes-Wohlfarth criterion indicates a highly itinerant magnetism [13]. Regarding the latter, DSF=$\theta_{CW}/T_C$=1.05, implying a low degree of the spin frustration.

An analysis of the FC magnetization temperature dependence in the neighborhood of the magnetic transition with the critical scaling law $\chi \propto ((T-T_C)/T_C)^{-\gamma}$ is presented in Fig. 5. The most precise fit, carried out in the relative temperature range of $0.04 \leq (T-T_C)/T_C \leq 0.3$ for $H_{DC}$ = 50 Oe yielded the critical exponent $\gamma$ = 1.04 and $T_C$ = 167.5 K. The $\gamma$-value obtained is almost equal to $\gamma$ = 1, the value expected from the mean field theory, while $T_C$ conforms to the temperature of the $\chi''$ inflection point as well as to the temperature of the $dM_{REM}/dT$ maximum (not shown in this paper). It also compares well with the one determined with the inflection method.

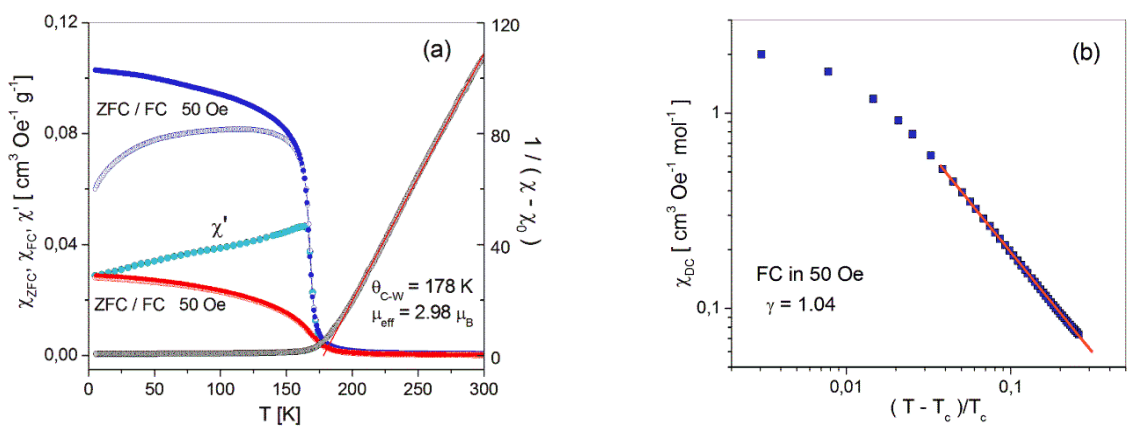

Fig. 4 (a) Static magnetic susceptibilities measured in the ZFC and FC modes in the DC fields of 50 Oe and 1 kOe. A fit of the Curie-Weiss law to the FC data in the paramagnetic range is here also shown, and (b) determination of the critical exponent $\gamma$ from the $\chi_{FC}$ data measured at $H_{DC}$=50 Oe.

A series of FC magnetization curves measured vs. temperature in different external magnetic fields ranging from 20 Oe to 2 kOe is displayed in Fig. 5a. The aim of these measurements was to trace the dependence of $T_C$ on magnetic field in order to see whether or not singularities observed in the $\chi'$ vs. $T$ curves (Fig. 3) are also reflected in the DC magnetization curves. As shown in Fig. 5b, the $dM_{FC}/dT$ exhibits a deep that shifts to higher temperature with the field but, additionally, in the field $H \geq 200$ Oe, some structure appears, resembling to some degree the one observed in the $\chi'(T)$ curves. Moreover, the asymmetric shape of the $dM_{FC}/dT$ curve seen for $H = 20$ Oe is similar to that of the $d\chi''/dT$ minimum (Fig. 3b). The further shift of the transition temperature, in the fields up to 50 kOe, manifested in the magnetocaloric effect is described in the next section.

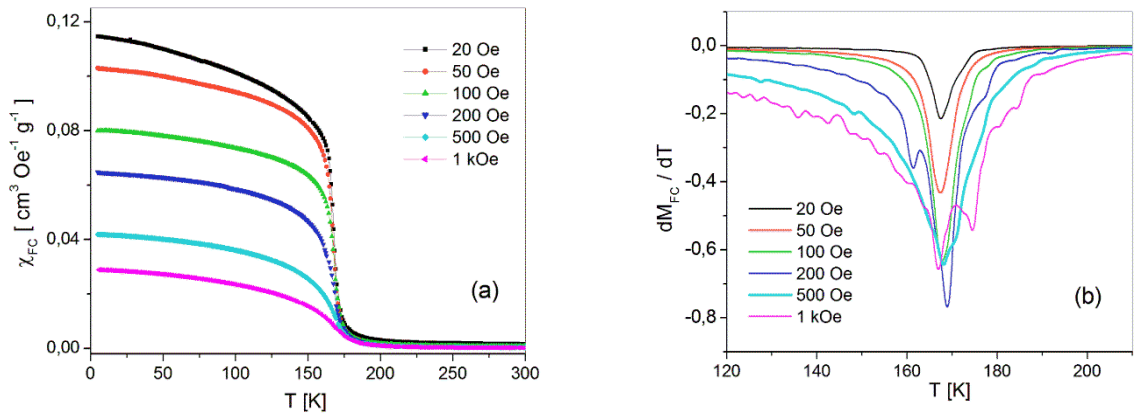

Fig. 5 (a) Temperature dependence of the FC susceptibility, $\chi_{FC}$, recorded in the magnetic field ranging between 20 Oe and 2 kOe, and (b) Temperature derivative of the FC magnetization curves, $dM_{FC}/dT$, as found for the $M_{FC}$-curves measured at different fields indicated.

Magnetization curves measured at $T = 2$ K, 60 K and 120 K for the sample cooled from 300 K in zero field condition are shown in Fig. 6. A spontaneous magnetic moment at $T = 2$ K is $\mu_S = 0.4\ \mu_B$, that is 7.5 times smaller than the effective moment (2.98 $\mu_B$) determined for the paramagnetic state. A $M(H)$ behavior with zero coercivity field is typical of a soft magnet. There is, however, a weak magnetic remanence present when the field is switched off after a cycle in which $H$ increases to 70 kOe and falls back to zero. At $T = 2$ K the isothermal remanence $M_{IRM} = 0.0064\ \mu_B$ mol$^{-1}$ (0.64 cm$^3$ g$^{-1}$), which is 1.6% of $M_S$. The exponential time dependence of $M_{IRM}$ can be described with two relaxation times $\tau_1$ and $\tau_2$, viz. $M_{IRM}(t)$ = constant + $A_1\exp(-t/\tau_1) + A_2\exp(-t/\tau_2)$. Fitting this dependence to the data shown in the inset of Fig. 6 gives $\tau_1 = (205 \pm 5)$ s, $\tau_2 = (2610 \pm 42)$ s, and $A_1/A_2 = 1.126$. While the value of $\tau_1$ is comprehensible for a soft magnet at $T = 2$ K, the long $\tau_2$ relaxation time ($\approx 40$ min) could be understood as associated with the reentrant spin-glass behavior and the slow misalignment of the ferromagnetic clusters after turn off of the field [18].

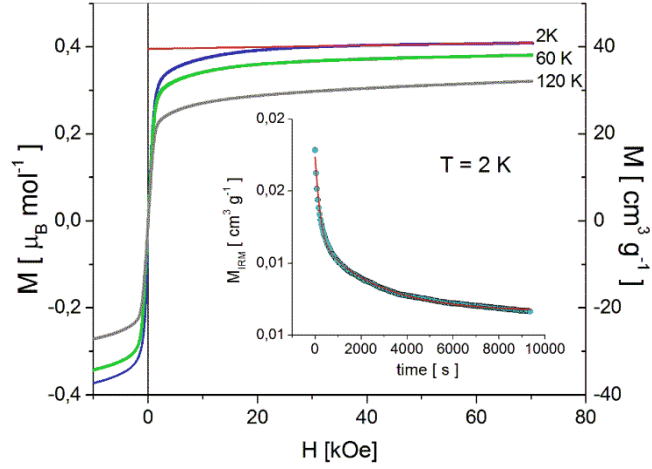

Fig. 6 Magnetization curves vs. magnetic field as measured at T = 2 K, 60 K and 120 K. Inset shows a decay of the isothermal magnetic remanence, $M_{IRM}$, with time.

## 4. Magnetocaloric effect

Here, magnetocaloric properties of the $\sigma$-$Fe_{60}V_{40}$ sample together with the data acquired for the previously investigated one viz. $\sigma$-$Fe_{66}V_{34}$ with $T_c$ = 312 K [4], which are included for comparison, are presented and discussed. The analysis of the results within the frame of the mean field approach is also performed.

The magnetic entropy change is given by the integrated Maxwell relation:

$$\Delta S(\Delta H, T) = \int_{H_0}^{H} \left( \frac{\partial M(T,H)}{\partial T} \right)_H dH \qquad (2)$$

where we put $H_0$ = 0. As follows from Eq. (2), $\Delta S$ is proportional to the temperature derivative of magnetization at constant field, which is the largest at the magnetic ordering temperature. In case when the specific heat $c_p$ is known, the adiabatic temperature change $\Delta T_{ad}$ may be obtained as $\Delta T_{ad} \approx -\frac{T \Delta S}{c_p}$. The effect for both samples was determined from the temperature dependences of the magnetization measured at the magnetic field in the range from 2.5 kOe up to 50 kOe (see Fig. 7a for $\sigma$-$Fe_{60}V_{40}$). Figure 7b shows a calculated magnetic entropy change at several values of the field change, $\Delta H$. It appears that, in comparison with the MCE data reported for materials containing rare earth metals or transition metal-based compounds of the first order transition, $\Delta S$ values reaching 1.40 J $K^{-1}$ $kg^{-1}$ at $\Delta H$ = 50 kOe are much smaller. However, the data obtained fall into the $\Delta S$ range reported for amorphous Fe-base alloys, for example those of the Fe-Zr-B type [19].

A shift of a temperature at which $\Delta S$ has a peak, $T_{peak}$, from the critical temperature of $T_c$ = 169 K to higher temperatures observed with increasing H is visible in Fig. 7b. As the $T_c$ - value was determined from the temperature derivative of $\chi'$ measured with $H_{AC}$ = 1 Oe, and in the DC data, the $T_c$ increase with the field was apparent (see Fig. 5b), the increase in $T_{peak}$ can be understood. It follows from the scaling laws of the magnetocaloric effect in second order phase transitions [20] that the $T_{peak}$-$T_c$ distance changes with the field as $(T_{peak}-T_c) \propto H^{1/\Delta}$, where $\Delta$ is one of the critical exponents. Figure 8a shows the log-log plot of the magnetic field dependence of the $\Delta S^{max}$ temperature shift relative to $T_c$, which yields the value $\Delta$ = 1.6±0.1 for the σ-$Fe_{60}V_{40}$ alloy, which is close to $\Delta$=1.5 predicted by the mean field theory. A similar plot for the σ-$Fe_{66}V_{34}$ sample with $T_c$ = 312 K [4] is presented in the same figure, while the $\Delta S$ vs. $T$ results determined at several $\Delta H$ values are displayed in Fig. 8b. The $\Delta$ exponent for σ-$Fe_{66}V_{34}$ is equal to 1.7±0.1. The maximum entropy change $\Delta S^{max}$ at $\Delta H$ = 50 kOe for the latter equals 1.86 J $K^{-1}$ $kg^{-1}$. This value is higher than the corresponding one (1.40 J $K^{-1}$ $kg^{-1}$) found for σ-$Fe_{60}V_{40}$. It remains, however, in line with the conclusion of Refs. [21, 22], where a decrease of $|\Delta S^{max}|$ with the magnetic ordering temperature $\left|\Delta S^{max}\right| \propto T_c^{-2/3}$ was predicted.

The relation concerns phase transitions of the second order for compounds of the same spin (i.e. spontaneous magnetic moment, $M_s$), while in σ-$Fe_{66}V_{34}$, due to the higher iron content than that in σ-$Fe_{60}V_{40}$, the $M_s$ moment in enhanced up to 0.6 $\mu_B$ $mol^{-1}$ [5], i. e. by 50% more than its value (0.4 $\mu_B$ $mol^{-1}$) determined presently for σ-$Fe_{60}V_{40}$ (see Fig. 9).

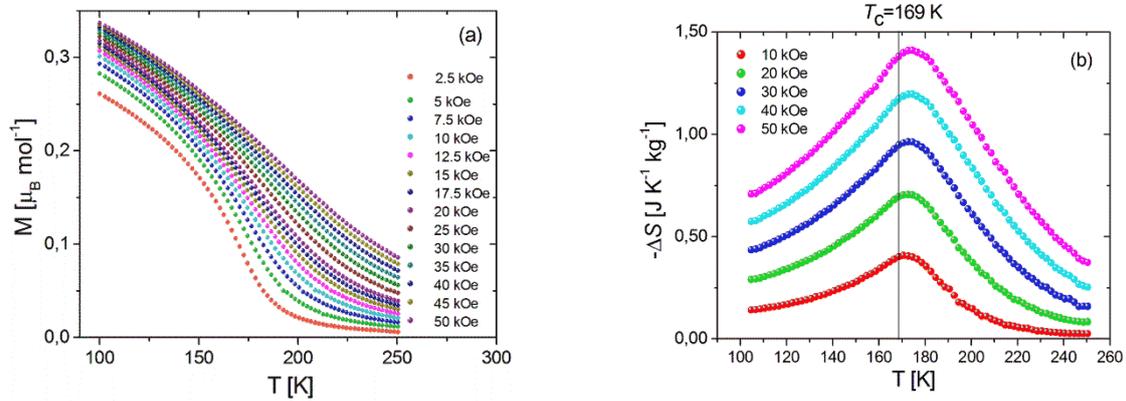

Fig. 7 (a) Temperature dependence of magnetization, $M$, for σ-$Fe_{60}V_{40}$ in different applied magnetic fields (shown as a legend), and (b) Magnetic entropy change, $\Delta S$, vs. temperature, $T$, for σ-$Fe_{60}V_{40}$ at different applied magnetic field change (shown as a legend).

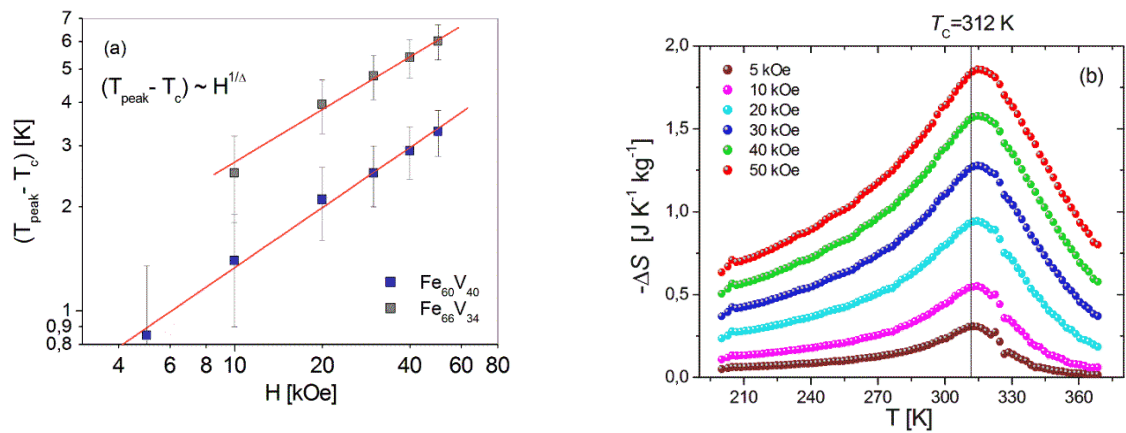

Fig. 8 (a) The log-log plot of the field dependence of the ($T_{peak}$-$T_c$) distance on the magnetic field, $H$, for σ-$Fe_{60}V_{40}$ and σ-$Fe_{66}V_{34}$, and (b) Magnetic entropy change, $\Delta S$, vs. temperature, $T$, for σ-$Fe_{66}V_{34}$ at different applied field change (shown as a legend).

Since the spin glass features observed in the σ-Fe$_{66}$V$_{34}$ and σ-Fe$_{60}$V$_{40}$ compounds are very weak, the mean field analysis of the magnetocaloric effect was performed. It was of interest to check whether or not $\Delta S(T)$ dependences on applied magnetic field may be presented as one master curve, like it is the case for ferromagnets. According to Ref. 22, the universal behaviour of $\Delta S$ curves at different applied fields is a hallmark of magnetic compounds with the second order phase transition, while for those with the first order phase transition, the breakdown of the universal curve is expected. Figures 9a and 9b present such phenomenological master curves constructed for σ-Fe$_{66}$V$_{34}$ and σ-Fe$_{60}$V$_{40}$, respectively. First, all $\Delta S(T)$ curves were normalized to their maximum values, $\Delta S^{max}$, and then the temperature axis below and above critical temperature was rescaled to get a new variable, $\theta$, where

$$\theta = \begin{cases} -(T - T_c)/(T_1 - T_c) & T \leq T_c, \\ (T - T_c)/(T_2 - T_c) & T > T_c, \end{cases} \quad (3)$$

Here $T_1$ and $T_2$ are temperatures of two reference points, and $h$ is a level of reference $\Delta S(T_1)/\Delta S^{max} = \Delta S(T_2)/\Delta S^{max} = h,$ where $h = 0.6$. It follows from Figs. 9a and 9b that for both σ phases investigated in this study the paramagnetic part of the $\Delta S(T)/\Delta S(T)^{max}$ curves collapses nicely onto a single curve, while at temperatures starting slightly below $T_c$, the universal behaviour does not occur. Obviously, the reason for this discrepancy is the re-entrant SG state observed in both alloys.

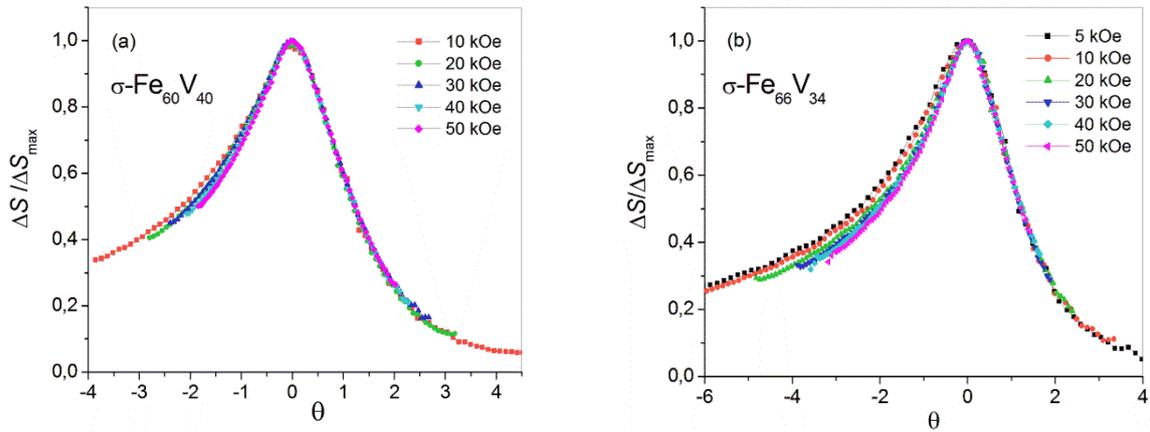

Fig. 9 (a) Relative change of the entropy vs. parameter θ for σ-Fe$_{60}$V$_{40}$ obtained for measurements in various fields (shown as a legend), and (b) the same for σ-Fe$_{66}$V$_{34}$.

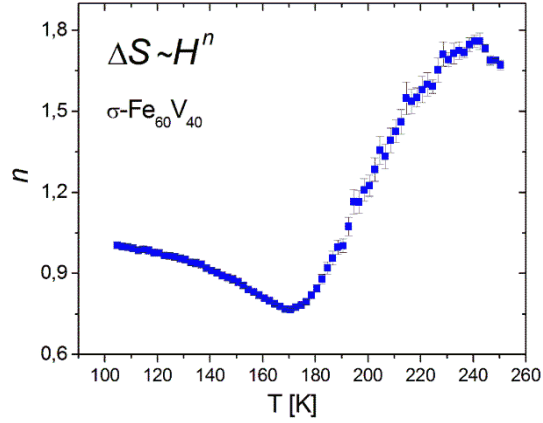

Fig. 10 Temperature dependence of the average exponent *n* for σ-Fe$_{60}$V$_{40}$ based on the data from Fig. 9a.

The mean field theory (MFT) predicts also that for materials with a second - order phase transition Δ*S* is proportional to $H^n$, where the critical exponent, *n*, is temperature dependent: for *T* well below $T_c$ *n*=1, at *T* =$T_c$ *n*=2/3, and *n*= 2 well above $T_c$ [20, 22]. A temperature dependence of the average value of *n* obtained for the σ-Fe$_{60}$V$_{40}$ alloy is presented in Fig. 10. It is clear that at *T*≈170 K, *n* has its minimum value of 0.75, which is larger than the value expected from the mean field theory, yet close to *n(T$_c$)* = 0.73 reported for an Fe-Zr-B-Cu amorphous alloy [22]. Taking into account relationships between various critical exponents given by scaling power laws, one can use Δ and *n(T$_c$)* values determined in the present study to estimate other critical exponents for σ-Fe$_{60}$V$_{40}$. Thus, using the following relations: Δ = β+γ and *n(T$_c$)* = 1+(β-1)/(β+γ), one receives the *β* exponent (relevant to a temperature dependence of the spontaneous magnetization) equal to 0.6, and γ exponent equal to 1.0, both in a good agreement with the MFT predictions. Noteworthy, for the σ-Fe$_{52}$V$_{48}$ compound ($T_C$=42K) β=1.0 and γ=1.55 were determined based on DC measurements and using conventional or extended scaling protocols [2]. Recent DC magnetic susceptibility study performed on σ-Fe$_{66}$V$_{34}$ and σ-Fe$_{64.5}$V$_{35.5}$ samples gave γ=1.4 (at 130 Oe) for the former and γ=1.3 (at 30 Oe) for the latter [3].

5. Conclusions

The results of this study permitted characterisation of the magnetic properties of the σ-Fe$_{60}$V$_{40}$ alloy as well as getting information on the magnetic universality class of the investigated sample. In particular, the following conclusions can be drawn:

1. The σ-Fe$_{60}$V$_{40}$ alloy shows a re-entrant behaviour with the ferromagnetic ordering temperature $T_C$ ≈169 K and the spin-freezing one $T_f$≈164 K.

2. The spin-freezing temperature exhibits a very weak increase with frequency: the relevant figure of merit being 0.002.

3. The degree of spin frustration is equal to 1.05.

4. Analysis of the magnetocaloric effect data yielded critical exponents $\Delta$=1.6 and $n(T_C)$=0.75.

5. Based on the relationships between various critical exponents given by scaling power laws and using $\Delta$ and $n(T_C)$ values obtained in this study, the critical exponents $\beta$=0.6 and $\gamma$=1.0 were determined.